\documentclass{icrc}

\usepackage{times}
\usepackage{graphicx} % when using Latex and dvips
\begin{document}

\title{Search for point sources and diffuse emission from
the Galactic plane with the HEGRA-IACT-system}
\author[]{H. Lampeitl}
\author[]{K. Bernl\"ohr}
\author[]{A. Daum}
\author[]{W. Hofmann}
\author[]{A. Konopelko}
\author[]{G. P\"uhlhofer}
\affil[]{Max Planck Institut f\"ur Kernphysik, Postfach 103980, D-69029 Heidelberg}
\author[ ]{HEGRA-Collaboration}

\correspondence{H. Lampeitl \\(Email: lampeitl@daniel.mpi-hd.mpg.de)}

\firstpage{1}
\pubyear{2001}

% \titleheight{11cm} % uncomment and adjust in case your title block
                     % does not fit into the default and minimum 7.5 cm

\maketitle
\begin{abstract}
The HEGRA-IACT-system with a FoV of $\sim 1.5$ deg radius has been
used for
surveying one quater of the  Galactic disc in respect to point
sources, moderately extended sources
and for diffuse emission in the energy range above 1 TeV. In total
140 h of good observation time were
accumulated. No new source has been discoverd. Limits on
the level of 20\% or lower of the Crab flux on about 87 potential
sources like SNR, Pulsars and EGRET sources are
derived. A limit on the diffuse emission is given on the level of
$d\Phi/dE$(E= 1 TeV) = 6.1 $\cdot$ 10$^{-15}$ ph cm$^{-2}$ s$^{-1}$
sr$^{-1}$ Mev$^{-1}$
resulting in a lower limit of 2.5 on the spectral index for the
extrapolation of the meassured EGRET flux for the diffuse emission.
\end{abstract}

\section{Introduction}
\begin{figure*}[t] 
\includegraphics[width=16.4cm]{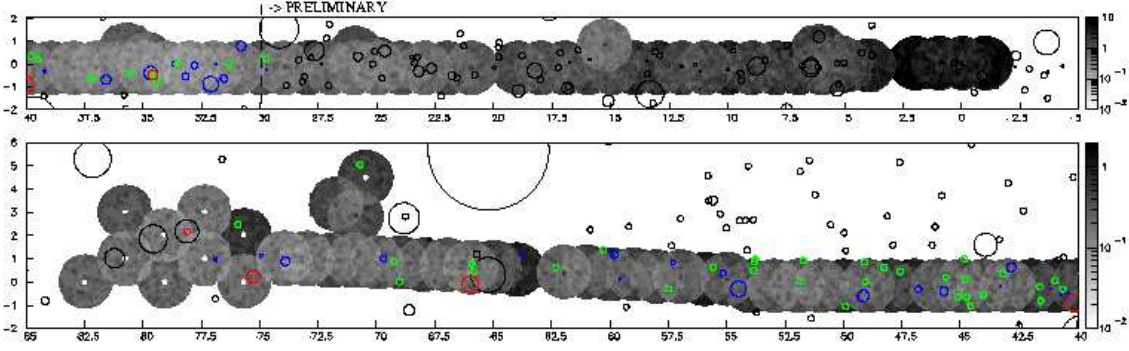} 
\caption{Galactic Longitude Scan (GLP). Derived upper limits for point sources
in units of the flux from the Crab Nebula. Circles indicate positions of
potential TeV-gamma-ray emitters like Pulsars, SNR and EGRET GeV sources.} 
\label{glp_scan}
\end{figure*}
\begin{figure}[ht] 
\vspace*{2.0mm} % just in case for shifting the figure slightly down 
\includegraphics[width=6.0cm]{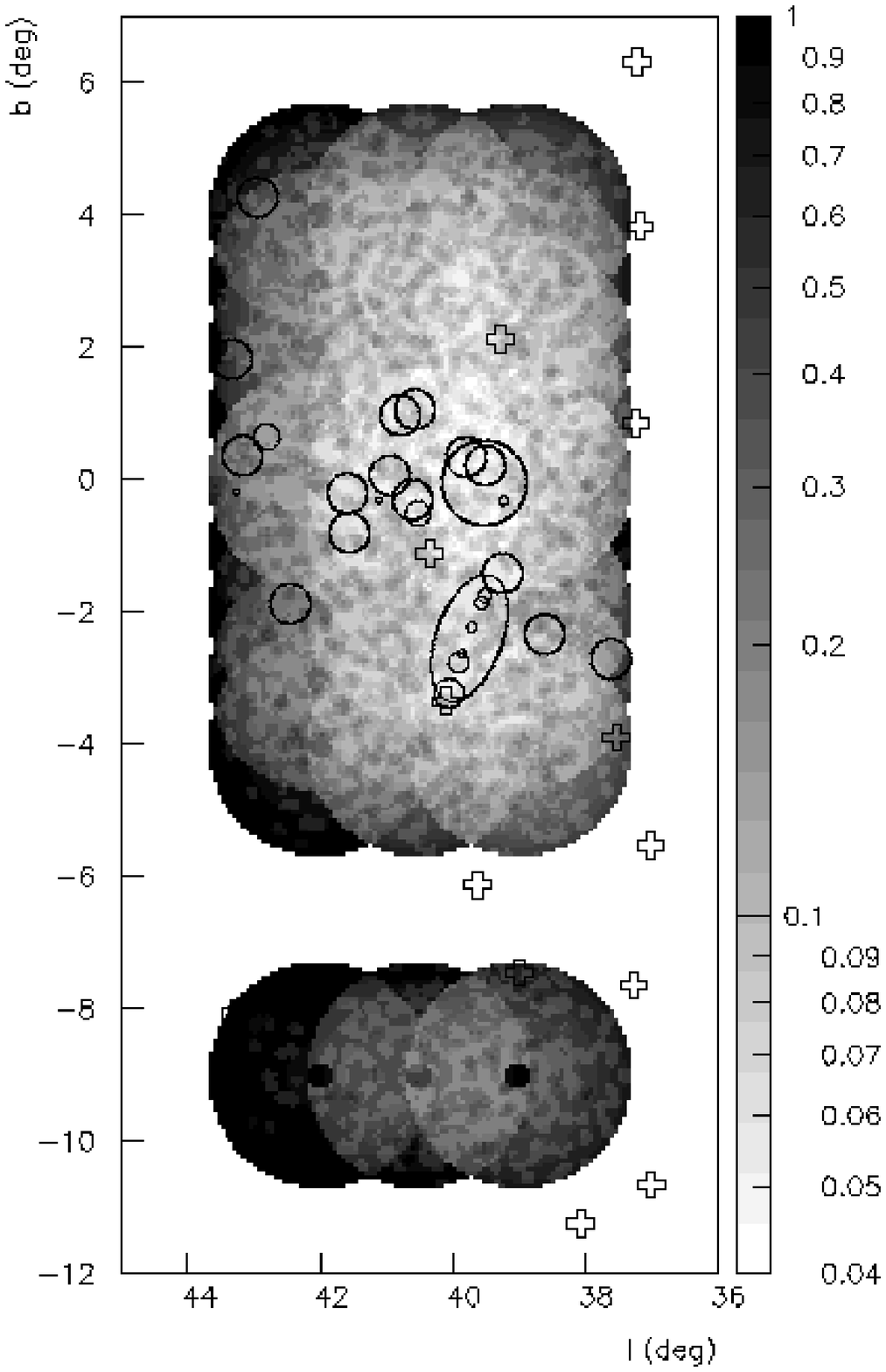}   % .eps for Latex, 
                                           % pdfLatex allows .pdf, .jpg, .png and .tif 
\caption{Galactic Latitude Scan (GLS). Derived upper limits for point sources
in units of the flux from the Crab Nebula.}
\label{gls_scan}
\end{figure}

 
Systems of imaging atmospheric Cherenkov telescopes, such as the
HEGRA stereoscopic telescope system 
(Daum et al. 1997, Konopelko et al. 1999), allow to reconstruct
the directions of air showers over the full field of view,
with a radius of about 1.5$^\circ$, and can therefore be used for
sky surveys. 
Here, we report on two surveys, one along the galactic equator 
(GLP, Fig. \ref{glp_scan}) with the focus on the search for point sources  
and a second one perpendicular to the galactic equator 
(GLS, Fig. \ref{gls_scan}) with the focus on diffuse emission. 
The GLP scan, covering one quarter of 
the Galactic disk, from the galactic center to the Cygnus region,
includes 105 h of observation time and is described in detail 
in P\"uhlhofer et al., 1999. Here we give results on individual 
point sources.\\
The second scan (GLS) covers a rectangular
patch of the sky of roughly 80~deg$^2$, centered on the
Galactic plane at longitude $40^\circ$ and includes 42 h of observation.
The motivation for this survey was mainly the
search for diffuse gamma-ray emission from the Galactic
plane.\\
%\item Search for gamma-ray point sources
%\end{itemize}
%
Diffuse emission results from the
interactions of charged cosmic rays
with interstellar gas confined to the plane or with photons.
Diffuse emission
in the energy range from tens of MeV to tens of GeV has been
studied recently by
% SAS~2 (Fichtel et al. 1975; Hartmann et al. 1979),
% COS~B (Mayer-Hasselwander et al. 1980, 1982) and 
EGRET (Hunter et al. 1997). The basic features can be modeled assuming
$\pi^o$ decay as the dominant mechanism, with the gamma-ray emission
proportional to the product of the gas column density and
the cosmic-ray density (see, e.g., Hunter et al. 1997).
%The distribution of cosmic rays is assumed to follow the matter density
%with a characteristic correlation scale of 1.5 to 2 kpc
%(Hunter et al. 1997).
%In addition to diffuse Galactic gamma rays, there is also a small
%extragalactic component, which should be fairly isotropic
%There is also evidence for a small isotropic extragalactic component
%(Fichtel et al. 1978; Sreekumar et al. 1998). 
Above 1 GeV, data show an
excess in gamma-ray flux over model predictions
(Hunter et al. 1997; see however Aharonian and Atoyan 2000). At these
energies, contributions from inverse Compton scattering
of electrons start to become relevant
(see, e.g., Porter \& Protheroe 1997). In response,
revised models speculate that the local measurements of
the electron flux may not be representative for the entire
Galaxy (Porter \& Protheroe 1997; Pohl \& Esposito 1998).
%
%Electron propagation is limited by radiative
%losses, and the local electron spectra are strongly influenced
%by the history of sources in the local neighborhood
%(Aharonian 1995a; Pohl \& Esposito 1998).
In case that the solar system is in an ``electron void'',
diffuse gamma-ray emission at high energies could be an
order of magnitude above predictions based on local electron
spectra (Pohl \& Esposito 1998).
%Detailed models of the full spectrum of diffuse gamma-ray
%emission from the Galaxy (see e.g. Moskalenko \& Strong 2000) 
%also favour a harder electron spectrum. 
Another 'diffuse'
gamma-ray flux component arises from the hard energy spectrum
of those Galactic CRs that are still confined in the ensemble
of their unresolved sources like SNRs (Berezhko \& V\"olk, 2000).\\ 
%
%Assuming these to be SNRs,
%Berezhko \& V\"olk (2000) showed that their spatially averaged
%contribution to the diffuse gamma-ray flux at 1~TeV should exceed
%the modell predictions of Hunter et al. (1997) by almost an order
%of magnitude.
%
%
%
\begin{table}[t]
\begin{center}
\begin{tabular}[h]{|l | r r ||  l | r r |}
\hline
Pulsar & $\sigma$ & [CU] & Pulsar & $\sigma$ & [CU] \\
\hline
%\multicolumn{6}{|l|}{Pulsars} \\ \hline
J1844-0244     & -1.0 & 0.28&   J1914+1122     &  0.8 & 0.23 \\    
J1848-0123     &  2.2 & 0.43&   J1915+07$^b$   & -1.1 & 0.22 \\  
J1852+00       & -0.3 & 0.14&   J1915+1009     & -0.7 & 0.14 \\  
J1854+10$^b$   & -0.8 & 0.29&   J1916+1030     & -1.7 & 0.1  \\  
J1856+0113     & -0.6 & 0.15&   J1916+0951     &  0.2 & 0.18 \\    
J1857+0057     & -0.8 & 0.13&   J1916+1312     &  1.0 & 0.34 \\  
J1857+0212     & -1.3 & 0.14&   J1917+1353     &  2.1 & 0.34 \\  
J1901+0331     & -0.9 & 0.18&   J1918+1444     &  0.4 & 0.32 \\  
J1901+0716$^b$ & 0.2  & 0.11&   J1921+1419     &  1.1 & 0.19 \\  
J1902+0556$^b$ & 0.5  & 0.06&   J1923+17       & -0.6 & 0.24 \\  
J1902+06$^b$   & -0.3 & 0.06&   J1926+1648     & -0.2 & 0.13 \\  
J1902+07$^b$   &  0.7 & 0.08&   J1926+1434     &  0.3 & 0.38 \\  
J1904+10$^b$   &  0   & 0.53&   J1927+1855     &  0   & 0.87 \\  
J1905+0709$^b$ & -0.2 & 0.08&   J1927+1856     & -0.5 & 0.36 \\  
J1906+0641$^b$ & -0.5 & 0.05&   J1929+18       & -0.7 & 0.13 \\  
J1908+07$^b$   & -0.7 & 0.07&   J1932+2020     & -1.1 & 0.18 \\               
J1908+0916     &  0.1 & 0.16&   J1939+24       & -0.3 & 0.21 \\  
J1908+04$^b$   & -2.4 & 0.05&   J1939+2134     &  0.6 & 0.17 \\  
J1908+0916$^b$ &  2.4 & 0.32&   J1946+26       &  0.4 & 0.19 \\  
J1909+1102     & -1.1 & 0.36&   J1948+3540     &1.7   & 0.56 \\  
J1909+0254$^b$ & -1.1 & 0.15&   J1954+2923     &0.3   & 0.13 \\  
J1910+0358$^b$ & -0.7 & 0.09&   J1955+2908     &-1.8  & 0.08 \\  
J1910+07$^b$   & 0    & 0.09&   J2002+3217     &-1.5  & 0.11 \\  
J1912+10       &  1.0 & 0.24&   J2004+3137     &-0.4  & 0.29 \\  
J1913+09       & -1.5 & 0.09&   J2013+3845     &0.4   & 0.46 \\  
\hline
\end{tabular}
\label{pointsearch44_ul}
\end{center}
\end{table}
\begin{table}[t]
\begin{center}
\begin{tabular}[h]{|l | r r ||  l | r r |}
\hline
SNR & $\sigma$ & [CU] & SNR & $\sigma$ & [CU] \\
\hline
%\multicolumn{6}{|l|}{SNR} \\ \hline
 G29.7-0.3    &-0.7 & 0.28  & G45.7-0.4$^3$ &-1.7 & 0.15 \\
 G30.7+1.0$^3$&-0.4 & 0.46  & G46.8-0.3$^2$ &1.0  & 0.35 \\
 G31.5-0.6$^2$&-0.9 & 0.19  & G49.2-0.7$^3$ &-0.5 & 0.21 \\
 G31.9+0.0$^1$&0.1  & 0.1   & G54.1+0.3     &-0.3 & 0.17 \\
 G32.1-0.9$^4$&0.8  & 0.6   & G54.4-0.3$^4$ &-0.6 & 0.27 \\
 G32.8-0.1$^2$&1.1  & 0.45  & G55.0+0.3$^2$ &-0.1 & 0.21 \\
 G33.2-0.6$^2$&1.4  & 0.25  & G57.2+0.8$^1$ &0.3  & 0.11 \\
 G33.6+0.1$^1$&0.4  & 0.15  & G59.5+0.1$^1$ &1.4  & 0.26 \\
 G34.7-0.4$^4$&-0.7 & 0.3   & G59.8+1.2$^2$ &0.6  & 0.4  \\
 G36.6-0.7$^3$&1.1  & 0.45  & G63.7+1.1$^1$ &0    & 0.35 \\
 G39.2-0.3$^{b}$&1.04& 0.08  & G69.7+1.0$^2$ &0.2  & 0.17 \\
 G40.5-0.5$^{2b}$&-0.9  & 0.06  & G73.9+0.9$^3$ &0    & 0.18 \\
 G41.1-0.3$^b$&1.0    & 0.1  & G74.9+1.2$^1$ &0.8  & 0.15 \\
 G42.8+0.6$^{2b}$&1.1  & 0.26   & G76.9+1.0$^1$ &0.4  & 0.13 \\
 G43.3-0.2    &-0.7 & 0.14  & & &\\
\hline
\hline
\multicolumn{6}{|l|}{GeV} \\ \hline
 J1856+0115$^3$& -2.7 & 0.11  & J2020+3658$^4$ & 0.4 & 0.27\\
 J1907+0557$^5$& -0.3 & 0.78  & J2020+4023$^2$ & 0   & 0.15\\
 J1957+2859$^5$& -2.3 & 0.2   &            &     &     \\
\hline
\end{tabular}
\caption{Significances and upper limits in units of the Crab flux (CU).
(b): Limit derived from GLS-Scan. Source bin radius:(1) $\vartheta=0.18^o$, 
(2) $\vartheta=0.2^o$, (3) $\vartheta=0.25^o$, $\vartheta=0.3^o$,
(4) $\vartheta=0.4^o$, (5) $\vartheta=0.5^o$. Source positions and
sizes taken from Tayler et al. 1993, Green 1998, Lamb et al. 1997.
Limits on W50 will be reported in Rowell G., 2001}
\label{point_sources}
\end{center}
\end{table}
%
%It is therefore of significant interest to search
%for the extension of the diffuse emission from the Galactic
%plane at higher energies. 
%Upper limits on diffuse gamma-ray
%emission at TeV energies have been reported by
%Reynolds et al. (1993), and LeBohec et al. (2000),
%at a level of a few $10^{-3}$ of the cosmic-ray flux.
%The most stringent limits at higher energies, above 100 TeV,
%were given by Borione et al. (1997), and constrain the diffuse
%flux to less than $3 \cdot 10^{-4}$ of the cosmic-ray flux.
%
The Galactic plane is also a region rich in potential
gamma-ray point
sources like SNRs as well as
pulsar driven nebulae and unidentified EGRET sources. 
% Both types of objects
%are almost certainly accelerators of cosmic rays and
%emitters of high-energy gamma radiation. 
Theoretical
models predict that typical gamma-ray fluxes from the majority
of these objects are below the detection threshold of the
current generation of instruments (see, e.g., Aharonian et al. 1997; 
Drury et al. 1994).
However, both the lack of knowledge of the individual source parameters as 
well as
approximations used in the modeling result in large
uncertainties in the predictions for individual objects, 
by an order of magnitude or more. 
%In addition to pulsars and supernova remnants, many
%unidentified EGRET sources lie in the Galactic plane. 
%Given the
%density of source objects, surveys provide an efficient
%way to search for gamma-ray emission.

% Citation either explicite in text as Abramovitz and Stegun (1964) 
% or in parenthesis (Abramovitz and Stegun, 1964; Aref, 1983). 
% Three authors as First et al. (year) or (First et al., year). 
% Alternatively use commands \citet{as64} and \citep{a83}. 
% 
%\newpage
%
\section{Point Sources}
Point sources were searched in both scans on a grid where the bin
position is taken to be the source position. A grid spacing 
below the angular resolution of $\approx 0.1^o$ is chosen
of $0.0625^o$ in case of the GLS-scan and of $0.03125^o$
in case of the GLP-scan. The radius of the source bin for point sources 
is chosen according the angular resolution of the telescope
configuration to optimize $S/\sqrt{B}$. In case of moderately extendet
objects the radius of the source bin was adapted such, that the complete
source fits in the search bin.
To estimate backgrounds, three
regions of the same size as the source region were used, 
rotated by 90$^\circ$, 180$^\circ$ and 270$^\circ$ around 
the telescope axis, relative to the source. Significances
for a detection are then calculated according to Li \& Ma (1983).
Since no high significant source bin is found
in both scans upper limits are calculated using the procedure of 
O. Helene (1983). To convert from counts to fluxes in the case of
the GLP-scan Crab data are used for calibration, in case of the 
GLS-scan the expected number of counts are derivied from MC 
simulations. The simulations show that the result 
on the differential flux upper limit depends only weak on the 
spectral index at 1 TeV (see Aharonian et al., 1995).\\
Fig. \ref{glp_scan} shows a map of resulting upper limits 
for point sources for the GLP-scan. Analysis is carried out
for sources with $l>30^o$ corresponding to zenith angels $< 30^o$
and results on individual sources are 
given in Tab. \ref{point_sources}.
Results on smaller galactic longitude are preliminary and no individual
upper limits are given. For the GLS-scan results are shown in 
Fig. \ref{gls_scan} and in Tab. \ref{point_sources}.
%\newpage

\section{Diffuse Emission} 
\begin{figure}[t] 
\vspace*{2.0mm} % just in case for shifting the figure slightly down 
\includegraphics[width=8.3cm]{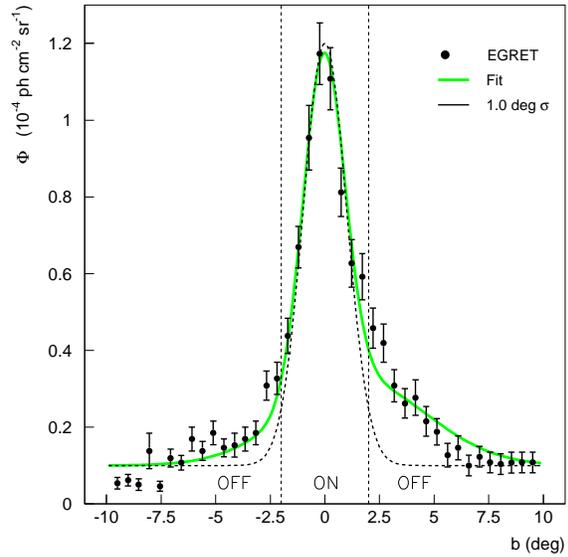} % .eps for Latex, 
%                                            % pdfLatex allows .pdf, .jpg, .png and .tif 
\caption{Latitude dependence of the diffuse emission measured by EGRET
for photon energies above 1 GeV,
in the range of Galactic longitude 30$^\circ$ to 50$^\circ$,
together with a fit by a sum of two Gaussians.
Data points taken from Hunter et al. 1997.
In addition a Gaussian profiles with a $\sigma$ of $1.0^\circ$,
is shown.}
\label{fig_difflat}
\end{figure} 
\begin{figure}[t] 
\vspace*{2.0mm} % just in case for shifting the figure slightly down 
\includegraphics[width=8.3cm]{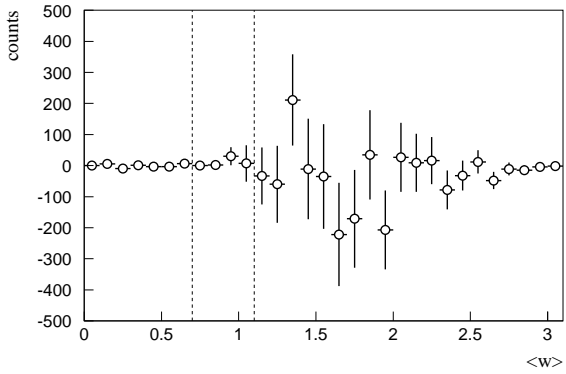} % .eps for Latex, 
%                                            % pdfLatex allows .pdf, .jpg, .png and .tif 
\caption{Difference between the distributions in mean scaled
width for the on-region ($|b| < 2^\circ$) and for the off-region
($|b| > 2^\circ$). The dashed lines indicate the expected gamma-ray
region.} 
\label{fig-diff-mscw}
\end{figure} 
\begin{figure}[t] 
\vspace*{2.0mm} % just in case for shifting the figure slightly down 
\includegraphics[width=8.3cm]{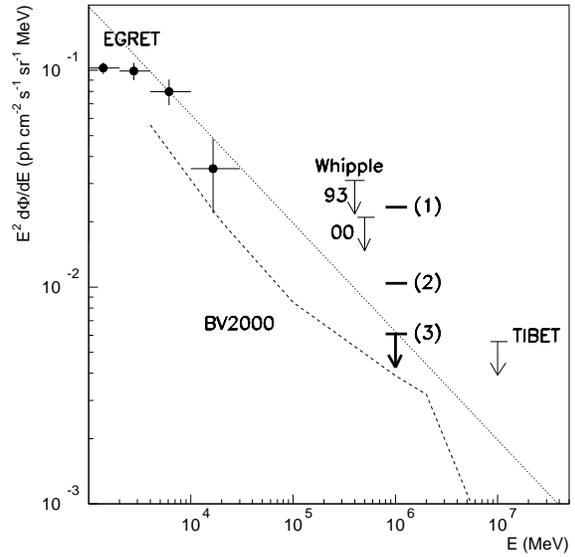} % .eps for Latex, 
%                                            % pdfLatex allows .pdf, .jpg, .png and .tif 
\caption{Upper limits for the diffuse gamma-ray flux derived
by this experiment. Assuming that all detected events are gamma-rays (1),
using a independent data set for background subtraction (2),
and using $|b|>2^\circ$ as background data and assuming the
spatial distribution measured by the ERGRET instrument
($38^\circ < l < 43^\circ, |b| < 2 ^\circ$) (3). For detailed explanations
see text.
Also shown is the EGRET flux for $35^\circ < l < 45^\circ,
|b| < 2^\circ$, the Whipple upper limits for
$38.5 < l < 41.5^\circ, |b| < 2^\circ$ (Reynolds et al. 1993, LeBohec 2000)
and the Tibet upper limit (Amenomori, 1997).
The dotted line indicates an extrapolation of the EGRET data
with an index of 2.5. The dashed line indicates the
scaled 'leaky box' model prediction by Berezhko \& V\"olk, 2000 (BV2000).} 
\label{fig-uplimits}
\end{figure} 
%
%Compared to the search for point sources,
The search for diffuse gamma-ray emission from the Galactic plane
is complicated by the extended structure of the emission region.
The structure may be extended in latitude beyond the field of view.
Fig.~\ref{fig_difflat} illustrates the profile in
Galactic latitude as measured by EGRET. \\
To achive low threshold, observations 
were carried out under small zenith angles ($z<30^o$) in June, July and August 1999.
Emphasis was on the control of systematic effects
like changing weather conditions.\\
The most robust and model-independent -- but also least sensitive --
technique to derive limits on the diffuse flux
simply selects 5-telescope-events according to their shapes
as gamma-ray candidates. A cut on $\bar{w}$ less than
1.0 keeps about 1/2 of the gamma-rays, but rejects cosmic rays very
efficiently.
%In order to achieve the best separation between gamma-ray
%images and cosmic-ray images, only five-telescope events were used.
%Such events with small $\bar{w}$ include genuine
%gamma-rays, electron showers from the diffuse cosmic-ray electron flux,
%and the tail of the distribution of cosmic-ray nuclear showers.
Assigning all 428 events after cuts as diffuse gamma-rays a 99\% upper
limit on the diffuse gamma-ray flux at 1 TeV of
23.4 $\cdot 10^{-15}$ ph cm$^{-2}$ s$^{-1}$ sr$^{-1}$ MeV$^{-1}$
results, for $|b| < 5^\circ$, assuming a flux from the Crab nebula
of $2.7\cdot10^{-17}$ ph cm$^{-2}$ s$^{-1}$ MeV$^{-1}$ at 1 TeV (Aharonian et al., 2000).
%The result depends in principle on the assumed spectral index. However,
%since flux values are quoted at energies corresponding roughly to the peak
%detection rates, the limits vary only very little with the spectral
%index. 
%We note that the diffuse
%electron flux (see, e.g., the compilation by Barwick et al. 1998)
%should contribute about 1/3 of the number
%of gamma-ray candidates; the limit can of course also be viewed
%as a limit on the electron flux, since the selection cuts are
%equally efficient for gamma-ray induced showers and electron-induced
%showers.
\\
The limit obtained by this technique can be improved by subtracting,
on a statistical basis the background, using an
experimental background region, sufficiently far away from the Galactic plane.
%In order to minimize instrumental effects, the background sample
%should be taken at the same time, and at identical zenith angles.
Data sets can be normalized to each other using the rates of events with large
$\bar{w}$ ($> 1.4$), well outside the gamma-ray region.
Unfortunately, the availability of suitable background data samples
with the same telescope configuration
is limited to 4.1 h of data, limiting the sensitivity.
After a $\bar{w}<1.1$ cut 1928 events survived compared
to 141 events in the reference sample. 
%The scaling factor between the two datasets is 
%determined to 13.3 and the MC simulations predict 555 events for a Crab like
%source smeared out over a FoV of $1.5^\circ$.
After subtraction of isotropic components, we find a 99\% limit on the
diffuse flux in the region $|b| < 5^\circ$ of
10.4 $\cdot 10^{-15}$ ph cm$^{-2}$ s$^{-1}$ sr$^{-1}$ MeV$^{-1}$ at 1 TeV.
\\
The final, and most sensitive analysis makes the assumption that
diffuse gamma-ray emission from the Galactic plane is limited
to the central parts of the scan region, and uses the outer
parts of the scan region to estimate backgrounds.
The signal region was considered  $|b| < 2^\circ$.
%This cut is close to optimal for emission profiles with an
%rms width between 1$^\circ$ and 2$^\circ$ and {\bf results in
%a good balance in observation time
%($\alpha \approx 1$) between the ON and OFF regions}.
To ensure optimum quality of the events, only four- and five-telescope
events were used, and the field of view was restricted to
1.5$^\circ$ from the optical axis. A cut at 1.1 on the
$\bar{w}$ was applied to reject cosmic-ray background.
To account for a possible zenith-angle dependence of
background rates, data were grouped into four different
ranges in zenith angle, 20$^\circ$-24$^\circ$, 24$^\circ$-28$^\circ$,
28$^\circ$-32$^\circ$
and 32$^\circ$-36$^\circ$. For each range in zenith angle and
each scan band, the expected number of events in the signal
region was calculated based on the number of events observed
in the corresponding areas of the camera for the background
region. The expected and observed numbers of events were then
added up for all zenith angles and scan bands.
With a total number of 2387 gamma-ray events in the signal region,
compared to 2353 expected events, there is no significant
excess (see Fig. \ref{fig-diff-mscw}).\\
In order to translate the limit in the number of events into
a flux limit, one now has to make assumptions concerning the
distribution in Galactic latitude of the diffuse radiation,
since a spill-over of diffuse gamma-rays into the background
region $|b| > 2^\circ$ will effectively reduce the signal.
For a profile with a width less or comparable to the
EGRET profile a correction of 12\% is applied
and one finds a limit
$6.1\cdot10^{-15}$ ph cm$^{-2}$ s$^{-1}$ sr$^{-1}$ MeV$^{-1}$ for the
diffuse gamma-ray flux at 1 TeV, averaged over the $|b| < 2^\circ$ region.
The limit
refers to an assumed spectral index of -2.6, and changes
by +13\% for an index of -2, and by -5\% for an index of -3.
For wider distributions of $2^\circ$ and $3^\circ$ rms, the
limit changes to $8.2 \cdot10^{-15}$ ph cm$^{-2}$ s$^{-1}$ sr$^{-1}$ MeV$^{-1}$
and $12.8\cdot10^{-15}$ ph cm$^{-2}$ s$^{-1}$ sr$^{-1}$ MeV$^{-1}$,
respectively. Results are summarized in Fig. \ref{fig-uplimits}.
% Vectors should be in bold italics with command $\vec{a}$, matrices 
% in bold roman with command $\mathbf{A}$, both in math mode. Units 
% should be in roman, either outside the math mode or with 
% $\mathrm{cm}$ in math mode. 
 
% Three typs of figure inclusion are demonstrated. Comment out the 
% appropriate commands. 
% (i) one column figure, will be floated to top of next column 
% 
% \begin{figure}[t] 
% \vspace*{2.0mm} % just in case for shifting the figure slightly down 
% \includegraphics[width=8.3cm]{figfile.eps} % .eps for Latex, 
%                                            % pdfLatex allows .pdf, .jpg, .png and .tif 
% \caption{Figure caption text.} 
% \end{figure} 
 
% (ii) two column figure, will be floated to top of next page 
% 
% \begin{figure*}[t] 
% \includegraphics[width=17.0cm]{figfile.eps} 
% \caption{Figure caption text.} 
% \end{figure*} 
 
% (iii) 1 1/2 column figure with caption on the side, will be floated 
% to top of next page 
% \begin{figure*}[t] 
% \figbox*{}{}{\includegraphics*[width=11.0cm]{figfile.eps}} 
% \caption{Figure caption text.} 
% \end{figure*} 
 
%\section{Balancing} 
 
% The columns of the last page can be balanced either by using the 
% command \verb/\balance/ somewhere in the first column of the last page 
% or by explicitely put \verb/\vadjust{\newpage}/ at the correct place. 
% without the \verb/ / command 
 
\section{Concluding remarks}

No strong TeV-gamma-ray emitter is detected in roughly one quater of the
Galactic disc coverd by both scans. Limits derived for known sources are 
on the level or lower than 20\% of the Crab flux. \\
A search for diffuse gamma-ray emission resulted in
an upper limit of $6.1\cdot10^{-15}$ ph cm$^{-2}$ s$^{-1}$ sr$^{-1}$ MeV$^{-1}$ at 1 TeV
averaged over the region $38^\circ < l < 43^\circ, |b| < 2 ^\circ$
and assuming the spatial emission profile measured by the EGRET instrument.
Since the analysis used to derive this limit
is only sensitive to the variation
of the diffuse flux with $b$, rather than its absolute value,
a distribution significantly wider than at EGRET energies
will increase the limit. Other variants of the data
analysis are sensitive to the absolute flux, but
give less stringent limits of $23.4$ and
$10.4\cdot10^{-15}$ ph cm$^{-2}$ s$^{-1}$ sr$^{-1}$ MeV$^{-1}$.
The limit on the TeV gamma-ray flux can be used to
derive a lower limit on the spectral index of the
diffuse radiation of 2.5, and to exclude models
which predict a strong enhancement of the diffuse
flux.
%compared to conventional mechanisms.
%However the TeV flux limit is only about a factor of 1.5 larger than the
%predicted flux from unresolved SNRs. A more sensitive survey should therefore
%be able to test this prediction, together with a determination of the
%"diffuse" TeV gamma-ray spectrum that is directly related to the Galactic
%CR source spectrum.

\begin{acknowledgements} 
The support of the HEGRA experiment by the German Ministry for Research
and Technology BMBF and by the Spanish Research Council
CYCIT is acknowledged. 
%We are grateful to the Instituto
%de Astrof\'\i sica de Canarias for the use of the site and
%for providing excellent working conditions. 
%We gratefully
%acknowledge the technical support staff of Heidelberg,
%Kiel, Munich, and Yerevan.
\end{acknowledgements} 
 
%\appendix 
%\section{Appendix section 1} 
%Text in appendix. 

\end{document}